\documentclass[preprint2]{aastex}
\usepackage{color}

\shorttitle{Coronal heating model} \shortauthors{Tan}

\begin{document}

\title{Coronal Heating Driven by A Magnetic Gradient Pumping Mechanism in Solar Plasmas}

\author{Baolin Tan}
\affil{Key Laboratory of Solar Activity, National Astronomical
Observatories of Chinese Academy of Sciences, Datun Road 20A,
Chaoyang District, Beijing 100012, China; bltan@nao.cas.cn}

\begin{abstract}

The heating of the solar coronal is a longstanding mystery in
astrophysics. Considering that the solar magnetic field is spatially
inhomogeneous with a considerable magnetic gradient from the solar
surface to the corona, this work proposes a magnetic gradient
pumping (MGP) mechanism to try to explain the formation of hot
plasma upflows, such as hot type II spicules and hot plasma
ejections. In the MGP mechanism, the magnetic gradient may drive the
energetic particles to move upward from the underlying solar
atmosphere and form hot upflows. These upflow energetic particles
are deposited in the corona, causing it to become very hot. Rough
estimations indicate that the solar corona can be heated to above 1
million degrees, and the upflow velocity is about 40 km s$^{-1}$ in
the chromosphere and about 130 km s$^{-1}$ in the corona. The solar
magnetic flux tubes act as pumpers to extract energetic particles
from the underlying thermal photosphere, convey them, and deposit
them in the corona. The deposit of these energetic particles causes
the corona to become hot, and the escape of such particles from the
photosphere leaves it a bit cold. This mechanism can present a
natural explanation to the mystery of solar coronal heating.

\end{abstract}

Online-only material: color figures

\section{Introduction}

As early as the 1940's, it was known that the solar corona is hot
enough that the temperature can exceed one million degrees, which is
about two orders of magnitude hotter than the temperature in the
underlying photosphere (Edlen 1943). It is believed that the energy
heating the corona comes from the solar interior. We also believe
that the solar magnetic field must play a key role in heating and
sustaining the hot corona. As for the heating power requirement
($P_{H}$), Aschwanden et al. (2007) estimate about
2$\times10^{5}\leq P_{H}\leq$2$\times10^{6}$ erg cm$^{-2}$s$^{-1}$
in active regions, about 1$\times10^{4}\leq
P_{H}\leq$2$\times10^{5}$ erg cm$^{-2}$s$^{-1}$ in quiet-Sun
regions, and about 5$\times10^{3}\leq P_{H}\leq$1$\times10^{4}$ erg
cm$^{-2}$s$^{-1}$ in coronal holes. There is still debate over how
the magnetic field transports the energy from the solar interior to
the outer atmosphere and how the energy is deposited once it reaches
the corona. According to thermodynamic laws, if only the thermal
conduction mechanism is at work, the temperature must steadily drop
from the photosphere to the corona with increasing height. The
heating mechanism of the extreme hot corona has puzzled
astrophysicists and theoreticians for more than 70 years. To solve
this big mystery, we must answer three key problems. (1) What is the
energy source? (2) How is the energy transported from the source
region into the corona? (3) How does the energy dissipate into heat?

A number of models have been proposed to solve this mystery.
Comprehensive reviews of this issue can be found in Narain \&
Ulmschneider (1996), Walsh \& Ireland (2003), and Klimchuk (2006)
among others. These models can be classified into two types. (1)
Wave mechanisms in which the solar violent inner motions of the Sun
cause the magnetic field lines to oscillate, transporting energy
through the cold underlying atmosphere and depositing it into the
corona via magnetohydrodynamic wave dissipations (Davila 1987). The
type of wave most frequently mentioned in these models is Alfven
wave (Heyvaerts \& Priest 1983; Davila 1987; Wu \& Fang 2003; De
Pontieu et al. 2007; Jess et al. 2009; van Ballegooijen et al. 2011;
Chen \& Wu 2012; etc.). The key problem with such models is how
these waves dissipate their energy and heat the plasma in the
corona, although many works show that the waves can carry enough
energy to sustain the hot coronal temperatures (De Pontieu et al.
2007; Kerr 2012). (2) Reconnection mechanism in which the violent
inner motions of the Sun cause the magnetic field lines to twist and
trigger multiple magnetic reconnections with various scales in the
upper atmosphere, releasing energy in the form of nanoflares that
will heat and accelerate the coronal plasmas (Parker 1988; Sturrock
1999; Cargill \& Klimchuk 2004; Rappazzo et al. 2008; etc.).
However, the most unresolved problem with magnetic reconnection
mechanism models is whether nanoflares can carry enough energy to
heat the solar corona.

Recently, researchers conducting a series of observational
discoveries have proposed explanations for coronal heating including
hot plasma ejections along the ultrafine magnetic loop channels from
the solar surface upward to the corona (Ji et al. 2012), hot upflows
of type II spicules (De Pontieu et al. 2009, 2011), and rotating
magnetic networks such as magnetic tornadoes
(Wedemeter-B$\ddot{o}$hm et al. 2012) and EUV cyclones (Zhang \& Liu
2011). In particular, ubiquitous type II spicules, which are hot
upflows in fountain-like plasma jets that have been observed at
extreme ultraviolet (EUV) wavelengths by the Atmospheric Imaging
Assembly (AIA) on board NASA's \emph{Solar Dynamics Observatory}
(\emph{SDO}), can heat the plasma to hundreds of thousands of
degrees on their way to the corona, with a small fraction reaching a
million degrees. This can be considered a possible candidate for
explaining the coronal heating (De Pontieu et al. 2011). However,
there is no theory yet to explain how the hot upflows form or why
they become heated.

In this work, we propose that the ubiquitous magnetic gradient in
the solar atmosphere may provide a pumping mechanism that picks up
the fast and relatively high energy charged particles from the
underlying thermal plasmas, drives them to move upward, and deposits
them in corona, increasing the average kinetic energy and causing
the corona to become very hot. This new model is called the magnetic
gradient pumping (MGP) mechanism and may explain the several recent
discoveries noted above. In Section 2 we discuss the magnetic
configuration in the solar atmosphere, from the solar photosphere to
the corona. Section 3 presents the deduction of the MGP mechanism in
different magnetic configurations and their explanation of the above
recent discoveries. Finally, the conclusion and some discussions are
presented in Section 4.

\section{Magnetic Configuration in Solar Atmosphere}

It is well known that the solar magnetic field is highly
inhomogeneous and this inhomogeneity plays a dominant role in almost
all processes in the solar atmosphere. Since magnetic fields are
frozen in the coronal plasmas, soft X-ray or EUV imaging
observations may present the configuration of the coronal magnetic
field. Such observations show that the fundamental structure of the
solar magnetic field is magnetic flux tubes (Zwaan 1978), which can
be sorted into two types (Fisk \& Schwadron 2001; Aschwanden \&
Nightingale 2005; Litwin \& Rosner 1993): open magnetic flux tubes
and closed magnetic flux loops.

1. In open magnetic flux tubes, the magnetic field lines remain
attached to the solar photosphere and are dragged outward into the
higher corona or even stretched into the remote heliosphere (Fisk \&
Schwadron 2001). They may be associated with solar streamers, solar
winds, solar radio type III bursts (Reid \& Ratcliffe 2014), etc.

2. In closed magnetic flux loops, the field lines remain connected
with opposite magnetic polarities entirely attached to the
photosphere and form multi-scale loops and active regions. They are
possibly related to various solar eruptions (L\'{o}pez Fuentes et
al. 2006).

\begin{figure}[ht] 
\begin{center}
   \includegraphics[width=8.2 cm]{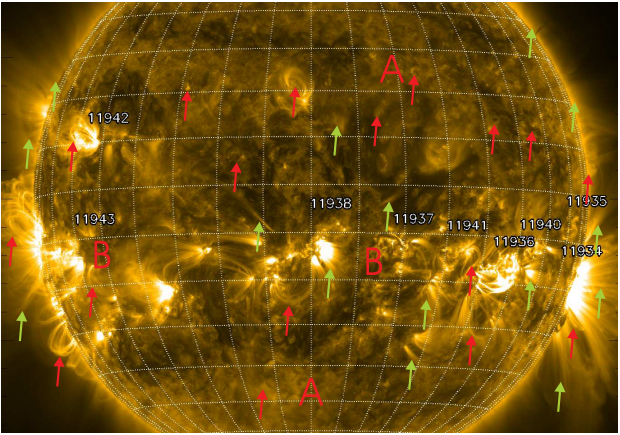}
\caption{EUV imaging observation of the solar disk at 171 \AA~ at
20:34:23 UT on 2014 January 1 obtained by AIA/SDO. Here, the open
magnetic flux tubes (green arrows) and the closed magnetic flux
loops (red arrows) can be found to have different spatial scales in
the quiet region (A) and active regions (B).}
\end{center}
\end{figure}

Practically, since the EUV line 171 \AA~ forms from Fe IX, its
formation temperature is about 6.3$\times10^{5}$ K, very closed to
the interface between the transition region and corona (Lemen et al.
2012). So, when a magnetic flux tube in the AIA 171 \AA~ image
extends deeply into the corona without a visible looptop, it can be
regarded as an open flux tube, while a magnetic flux tube connects
one footpoint to another via a visible looptop in the AIA 171 \AA~
image it can be regarded as a closed magnetic flux loop. Of course,
this classification is made relatively. Figure 1 presents a recent
imaging observation at 171 \AA~ observed by AIA/SDO with a pixel
size of 0$''$.6 on 2014 January 1. It is obvious that either in the
solar quiet regions (marked as $A$) or in the active regions (marked
as $B$), the magnetic configuration is composed of open magnetic
flux tubes (such as the place marked by green arrows) and closed
magnetic flux loops (such as the place marked by red arrows).
Actually, when we investigate a large closed magnetic flux loop,
such as big loops across different active regions, the local part
around one footpoint can be approximately regarded as an open
magnetic flux tube.

The common property of both types of configurations is that they
have a divergent structure with a considerable magnetic gradient
from the footpoints to the higher place. In fact, from imaging
observations, when we track a single coronal loop from one footpoint
via looptop to another footpoint, the loop's cross section is
approximately constant, but when we investigate a bundle of coronal
loops, we may find that they are diverging from their footpoints to
the higher corona. In such configurations, the magnetic gradients
are ubiquitous. Many researchers have tried to obtain the magnetic
gradients in solar atmosphere (Hagyard et al. 1983; Landolfi 1987;
Liu et al. 1996; Mathew \& Ambastha 2000; etc.). Gelfreikh et al.
(1997) obtained a coronal magnetic gradient of 10$^{-4}$ G km$^{-1}$
at a height of 10$^{5}$ km with a magnetic field strength of about
20 G. However, so far, it is still very difficult to obtain the
magnetic gradients in the solar atmosphere from observations.

\section{Magnetic Gradient Pumping Mechanism}

Schluter (1957) proposed that non-magnetized material bulk can be
pushed to move rapidly in an external diverging magnetic field as a
diamagnetic body and the acceleration is proportional to

\begin{equation}
-\nabla(logB^{2})
\end{equation}

Here $B$ is the external magnetic field. This idea is called the
melon-seed mechanism (Pneuman 1984). Many people adopted this
mechanism to explain the formation of solar surges, spicules, jets,
filaments, coronal mass ejections (CMEs), and solar wind expansion
(Altschuler et al. 1968; Pneuman 1983; Hollweg 2006; Filippov et al.
2007, 2009; Parashar et al. 2013). This mechanism can explain the
motion of the bulk of diamagnetic plasmas, but it cannot explain why
the corona is much hotter than the underlying chromosphere and
photosphere. Here, we investigate the kinetic behaviors of single
charged particles in open and closed solar magnetic flux tubes. We
find that different particles with different kinetic energies have
different motions.

\subsection{In Open Magnetic Flux Tubes}

Consider a charged particle moving in an open magnetic flux tube
with a converging region near the photosphere and a diverging region
in the corona and with a transverse velocity $v_{t}\neq 0$; it will
gyrate around the magnetic field lines in a circular or helical
orbit. The coupling of the radial component of the magnetic field
and the particle's transverse motion will produce an equivalent
driving force parallel to the magnetic field lines and pointing to a
weak magnetic field region (Figure 2). This force is called the
magnetic gradient force ($F_{m}$):

\begin{equation}      
F_{m}=-\mu \nabla B=-G_{B}\epsilon_{t}
\end{equation}

Here, $\mu=\frac{\epsilon_{t}}{B}$ is the particle's magnetic moment
which is approximately an invariance
$\epsilon_{k}=\epsilon_{t}+\epsilon_{l}$ where
$\epsilon_{t}=\frac{1}{2}mv_{t}^{2}$ is the transverse kinetic
energy, $\epsilon_{l}=\frac{1}{2}mv_{l}^{2}$ is the longitudinal
kinetic energy, $m$ is the particle's mass, and $v_{t}$ and $v_{l}$
are the transverse and longitudinal velocities, respectively.
$G_{B}=\frac{\nabla B}{B}$ is the relative magnetic gradient. The
absolute value of its reciprocal is the magnetic field scale height
$L_{B}=\frac{1}{|G_{B}|}=|\frac{B}{\nabla B}|$.

Equation (2) is very similar to Equation (1). However, they are
intrinsically different. Equation (1) only presents the magnetic
tension force acting on the bulk of diamagnetic materials in a
diverging field, while Equation (2) expresses the driving force
acting on a particle in a diverging magnetic field.

Here, there is another problem: Equation (2) works only in
collisionless plasmas where magnetic moment is conserved. The
photosphere, however, is highly collisional. As we know, the
collision frequency is proportional to the plasma density $n_{p}$,
and is inversely proportional to $T^{3/2}$. $T$ is the temperature.
Similar to the magnetic gradient, there is also a dominating density
gradient from the solar photosphere to the upper atmosphere
(chromosphere and corona), and the collision probability of a
particle will decrease rapidly from the underlying strong magnetic
field region to the upper weak field region. When a charged particle
moves from the dense photosphere to the tenuous corona, it is
reasonable to assume that the particle's magnetic moment is still
approximately conserved. Equation (2) is still approximately valid
in this regime.

In fact, $L_{B}$ changes with height ($h$) in the solar atmosphere.
It approximates to $\sim 1000$ km near the photosphere, $\sim 2000$
km in the chromosphere, and larger than $2\times10^{4}$ km in the
low corona (Gelfreikh et al. 1997; Verth et al. 2011). The magnetic
gradient in the solar open magnetic flux tube is downward, therefore
$F_{m}$ directs upward. The higher the transverse kinetic energy
($\epsilon_{t}$), the faster the particle moves upward. This leads
to separation of high-energy particles from lower energy particles.
When energetic particles are transported to the higher place and
reach a new thermal equilibrium through thermal collisions, they
reach a high temperature. As high-energy particles escape, the
average kinetic energy of the underlying adjacent plasma (e.g.,
around footpoints) will decrease and cool down slightly.

\begin{figure}[ht] 
\begin{center}
   \includegraphics[width=5 cm]{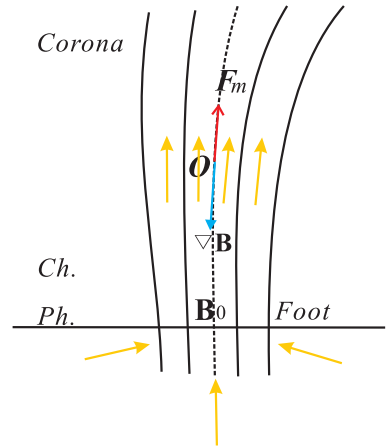}
\caption{Schematic diagram of an open magnetic flux tube. A charged
energetic particle at photosphere (Ph) will be driven to move upward
via the chromosphere (Ch) and into corona. $B_{0}$ is the magnetic
field at the footpoints and $d$ is the distance between footpoints.
The blue and red arrows indicate the magnetic gradient and its
driving force ($F_{m}$), respectively. Yellow arrows show
propagations of energetic particles.}
\end{center}
\end{figure}

In real magnetic flux tubes, besides the magnetic gradient force
$F_{m}$, the charged particles are also affected by the solar
gravitational force $F_{g}=mg(h)$. The net upward force acting on
particles can be expressed as

\begin{equation}    
f_{t}\simeq -G_{B}\epsilon_{t}-mg(h).
\end{equation}

Here, $g(h)$ is the solar gravitational acceleration at height $h$,
$g(0)\simeq 274$ m s$^{-2}$ near the photosphere. Only when
$f_{t}>0$ can the energetic particles be driven upward. Therefore, a
starting energy ($\epsilon_{t0}$) for the upward motion can be
deduced:

\begin{equation}     
\epsilon_{t0}= mg(h)L_{B}.
\end{equation}

Here we find that the starting energy is proportional to the mass of
the particle. Heavy ions have a higher starting energy and this
leads to a consequence that only the very high energy heavy
particles can be driven to move upward to the higher corona. For
example, Oxygen ions or metal ions will have several times or
decades of a proton's starting energy. In an open magnetic
configuration, the charged particles can be divided into two groups:

1. Confined particles, $\epsilon_{t}<\epsilon_{t0}$, which are
confined in the lower region of the magnetic configuration by the
solar gravitational force. Obviously, confined particles are
distributed in the lower energy part of the thermal distribution
profile.

2. Escaping particles, $\epsilon_{t}>\epsilon_{t0}$, which can be
driven to move upward along the magnetic flux tube. They are
distributed in the high-energy tail of the thermal distribution
function. Supposing that the solar photospheric plasma has a
Maxwellian distribution at temperature $T_{0}$, the plasma density
is $N_{0}$. The number of escaping particles can be calculated as

\begin{equation}   
N(\epsilon_{t}>\epsilon_{t0})=qN_{0}\int_{\epsilon_{t0}}^{\infty}
f(\epsilon_{k})d\epsilon_{k}
\end{equation}

Here, $f(\epsilon_{k})=2\pi[\frac{\epsilon_{k}}{ (\pi
k_{B}T_{0})^{3}}]^{\frac{1}{2}}e^{-\frac{\epsilon_{k}}{k_{B}T_{0}}}$,
$k_{B}$ is the Boltzman constant. $q$ is a factor indicating the
fraction of $\epsilon_{t}$ in total kinetic energy, $q\sim 0.5$ for
simplicity. The total energy carried by escaping particles is

\begin{equation}   
E(\epsilon_{t}>\epsilon_{t0})=qN_{0}\int_{\epsilon_{t0}}^{\infty}
f(\epsilon_{k})\epsilon_{k}d\epsilon_{k}
\end{equation}

The energetic particles escape from the underlying atmosphere, reach
a higher place, and reach thermal equilibrium by continuous
collisions. The particle's average kinetic energy can be regarded as
an estimation of the temperature:

\begin{equation}     
T_{c}=\frac{E(\epsilon_{t}>\epsilon_{t0})}{k_{B}N(\epsilon_{t}>\epsilon_{t0})}
\end{equation}

Assuming $L_{B}=1000$ km near the photosphere, the starting energy
$\epsilon_{t0}$ is 0.0016 eV for electrons and 2.85 eV for protons.
Here, the starting energy of electrons is much smaller than that of
ions. It seems that a great majority of electrons and only a small
part of ions can be driven to move upward. In fact, as the
electrostatic attraction between electrons and ions will hold back
electrons and drag ions upward to avoid the spatially charged
separation, ions will play a key role in the above regime. It is
reasonable to adopt the starting energy of ions as the lower limit
in the above integrating calculations. When
$\epsilon_{k}>\epsilon_{t0}$ and the charged particle moves from the
lower strong field region (e.g. near the photosphere) upward to the
upper weak field region (e.g., the corona), its transverse kinetic
energy will convert gradually into longitudinal kinetic energy. When
$\epsilon_{t}$ gets smaller and smaller and finally diminishes, the
particle approaches to a maximum longitudinal velocity. The higher
the initial total kinetic energy, the faster the particle moves
upward. The higher energy particles more easily reach a higher
region with fast longitudinal velocity and form a steady hot upflow.
The velocity of upflows can be estimated as

\begin{equation}
v_{up}=\frac{\int_{\epsilon_{t0}}^{\infty}
f(\epsilon_{k})\sqrt{\frac{2\epsilon_{k}}{m}}d\epsilon_{k}}{\int_{\epsilon_{t0}}^{\infty}
f(\epsilon_{k})d\epsilon_{k}}.
\end{equation}

The energy flow can be roughly estimated by

\begin{equation}
P_{up}\simeq k_{B}T_{c}\cdot N(\epsilon\geq\epsilon_{t0})\cdot
v_{up}
\end{equation}

When the underlying atmosphere loses energy
$E(\epsilon_{t}>\epsilon_{t0})$ to the upflow of escaping particles,
it drops to a low temperature:

\begin{equation}
T_{s}=\frac{N_{0}k_{B}T_{0}-E(\epsilon_{t}>\epsilon_{t0})}{k_{B}[N_{0}-N(\epsilon_{t}>\epsilon_{t0})]}
\end{equation}

The deficit of energetic particles in the thermal atmosphere can be
compensated for by the ambient and solar interior plasmas through
thermal diffusions and interior convection. The yellow arrows in
Figure 2 indicate the propagating direction of the energetic
particles from the solar interior to the corona.

Generally, near the solar photosphere, $N_{0}\simeq10^{14}$
cm$^{-3}$, $T_{0}\simeq 6000$ K (Vernazza et al. 1981). After the
filtration of magnetic gradient force, we may obtain: $T_{s}\sim
5933$ K, $N(\epsilon_{t}>\epsilon_{t0})\simeq5.88\times10^{11}$
cm$^{-3}$, and $T_{c}\sim3.94\times10^{4}$ K, $v_{up}\simeq25.5$ km
s$^{-1}$, and energy flow $P_{up}\simeq8.19\times10^{6}$ erg
cm$^{-2}s^{-1}$. These values are very close to that in the solar
chromosphere at about 2000 km height (Vernazza et al. 1981).

In the solar chromosphere, the magnetic gradient becomes
increasingly weaker and $L_{B}$ becomes increasingly longer than
that near the photosphere. Simply, we suppose that $L_{B}\sim 2000$
km, and the temperature is about $3.94\times10^{4}$ K. Then we
obtain $N(\epsilon_{t}>\epsilon_{t0})\simeq1.01\times10^{11}$
cm$^{-3}$, $T_{c}\sim1.12\times10^{5}$ K, $v_{up}\simeq42.4$
km$s^{-1}$, and the energy flow $P_{up}\simeq6.66\times10^{6}$ erg
cm$^{-2}s^{-1}$. These values are very close to that in the solar
transition region between the chromosphere and the corona at about
2000-3000 km in height.

In the solar transition region, supposing: $L_{B}\sim 5000$ km,
$T_{0}\sim1.12\times10^{5}$ K. Then
$N(\epsilon_{t}>\epsilon_{t0})\simeq2.03\times10^{10}$ cm$^{-3}$,
$T_{c}\sim 2.97\times10^{5}$ K, $v_{up}\simeq68.9$ km$s^{-1}$, and
energy flow $P_{up}\simeq5.76\times10^{6}$ erg cm$^{-2}s^{-1}$.
These parameters are very close to that in the upper part of the
solar transition region or at the bottom part of the solar corona.

Furthermore, near the bottom of the solar corona: $L_{B}\sim
3.0\times10^{4}$ km, $T_{0}\sim2.97\times10^{5}$ K. Then
$N(\epsilon_{t}>\epsilon_{t0})\simeq3.52\times10^{8}$ cm$^{-3}$,
$T_{c}\sim 1.06\times10^{6}$ K, $v_{up}\simeq132.9$ km$s^{-1}$, and
energy flow $P_{up}\simeq6.91\times10^{5}$ erg cm$^{-2}s^{-1}$.
These values are similar to the condition in the solar corona.

The above estimations indicate that the solar chromosphere and
corona can be heated by a cascading filtration driven by magnetic
gradient force. The estimated values of the upflow velocities are
very close to those of type II spicules observed by AIA/SDO and
Hinode (De Pontieu et al. 2011) and can present a reasonable
explanation of the formation of type II spicules and other hot
plasma upflows. The open magnetic flux tube acts as a pumper to
extract the energetic particles from the underlying thermal
photosphere and transport them to and deposit them in the corona.
This process can be called as the MGP mechanism.

Actually, the magnetic gradient force is an equivalent effect
parallel and opposite to the magnetic gradient $\nabla B$, which
acts on the guiding center and filtrates charged particles by their
kinetic energies. In these processes, the magnetic field can not do
any extra work on the charged particles; the total energy must be
conservative. When magnetic gradient force causes the longitudinal
kinetic energy ($\epsilon_{l}$) to increase, the transverse kinetic
energy ($\epsilon_{t}$) decreases simultaneously. The energy
conversion takes place between $\epsilon_{t}$ and $\epsilon_{l}$.
When $\epsilon_{t}$ is converted fully into $\epsilon_{l}$, the
magnetic gradient force disappears and a stable upflow forms.

In fact, charged energetic particles will not move upward endlessly.
Because the magnetic field becomes homogenous in the higher corona,
the magnetic gradient gradually vanishes ($G_{B}\sim 0$) and the
magnetic gradient force also fades away ($F_{m}\sim 0$). When the
condition $G_{B}<\frac{mg(h)}{\epsilon_{t}}$ is fulfilled, the net
upward force becomes negative, $f_{t}<0$. The particles are
decelerated by solar gravitational force.

With the MGP mechanism, the escaping particle's transverse kinetic
energy will convert into longitudinal kinetic energy, which will
result in a temperature anisotropy in the coronal plasmas; the ion's
longitudinal temperature ($T_{l}$) will be higher than the
transverse temperatures ($T_{t}$). This property may explain ion
temperature anisotropy in the solar winds (Hahn \& Savin 2013). The
temperature anisotropy triggers a series of plasma instabilities and
leads to secondary energy release and the formation of thermal
isotropic distributions.

\subsection{In Closed Magnetic Flux Loops}

Closed magnetic flux loops with various length scales are ubiquitous
in the solar chromosphere and corona. They connect one magnetic
polarity to the other in sunspots in active regions as well as
connecting one side to another in the magnetic network of granules
in solar-quiet regions (Priest et al. 2002). They are always
associated with many phenomena such as solar active regions and
flares, CMEs, (Aschwanden \& Nightingale 2005), and X-ray and EUV
bright points in the quiet Sun (Golub et al. 1977; Habbal \&
Withbroe 1981). Besides being located in active regions, closed
magnetic flux loops also exist in solar-quiet regions with
relatively small scales (Centeno et al. 2007). This property can
also be seen in Figure 1.

When a particle leaves one footpoint of the closed magnetic flux
loop, as the relative longitudinal magnetic gradient is downward
$G_{B}<0$, the particle will be driven to move upward by an upward
magnetic gradient force (Figure 3). Considering the solar
gravitational force $mg(h)$, the net force acting on the particle
can be expressed as

\begin{equation}
f_{t}=-G_{B}(h)\epsilon_{t}\sin\alpha-mg(h)
\end{equation}

Here, $\alpha$ is the angle between the magnetic field line and the
horizontal direction, which will be near zero around the looptop.
Particles can be activated to move upward only when $f_{t}>0$. Then
a starting energy can be deduced:

\begin{equation}
\epsilon_{t0}=mg(h)L_{B}\sin^{-1}\alpha
\end{equation}

\begin{figure}[ht] 
\begin{center}
   \includegraphics[width=8.0 cm]{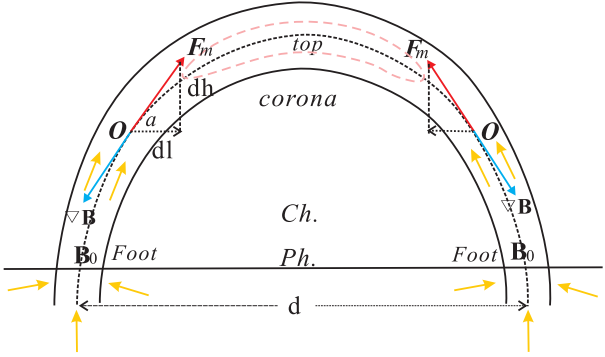}
\caption{Schematic diagram of closed magnetic flux loops. $B_{0}$ is
the magnetic field at the footpoints and $d$ is the distance between
footpoints. The blue and red arrows indicate the magnetic gradient
and its driving force ($F_{m}$), respectively. Yellow arrows show
the propagating flow of energetic particles. The dashed pink curve
presents the trajectory of bounce particles.}
\end{center}
\end{figure}

Supposing the magnetic field is approximately one dipole in closed
magnetic flux loops, it can be approximated as (Takakura \& Scalise
1970): $B(h)=(\frac{1}{1+h/d})^{3}B_{0}$. Here $d$ is the distance
between the two footpoints and $h$ is the height. Then
$\frac{\partial B}{B\partial l}=\frac{1}{B(h)}\frac{\partial
B}{\partial h}\frac{\partial h}{\partial
l}=-\frac{3}{d+h}\sin\alpha$, $L_{B}=(d+h)/3$. The magnetic gradient
diminishes around the looptop gradually. The starting energy of the
particles near the footpoint can be obtained as:
$\epsilon_{t0}\simeq \frac{1}{3}mdg(0)$. According to the particle's
kinetic energy ($\epsilon_{k}$) and the initial incident angle
($\theta$), all particles can be classified into three groups

1. \emph{Confined particles}. When the kinetic energy
$\epsilon_{k}<\epsilon_{t0}$, the particle is confined in the
underlying photospheric atmosphere. Of course, the confined
particles just have lower energy.

2. \emph{Passing particles}. When the kinetic energy
$\epsilon_{k}>\epsilon_{t0}$ and its initial incident angle
($\theta$) is smaller than the magnetic mirror critical angle
($\theta_{c}$), $\theta<\theta_{c}$, the particle can be driven to
move upward from one footpoint, pass the looptop and precipitate at
another footpoint. Here, the magnetic mirror critical angle
$\theta_{c}=arcsin\sqrt{1/R_{m}}$. $R_{m}=B_{max}/B_{min}$ is the
magnetic mirror ratio. $B_{max}$ and $B_{min}$ are the maximal and
minimal magnetic field strength in the loop, respectively.
Theoretically, the fraction of the passing particles is about
$(2R_{m})^{-1}$.

3. \emph{Bounce particles}. When the kinetic energy
$\epsilon_{k}>\epsilon_{t0}$ and its initial incident angle
$\theta>\theta_{c}$, the particle will be driven to move upward from
the footpoints, and bounce back and forth around the looptop (shown
as the dashed pink curve in Figure 3). The number of energetic
bounce particles can be estimated by

\begin{equation}
N_{b}=q_{m}qN_{0}\int_{\epsilon_{t0}}^{\infty}
f(\epsilon_{k})d\epsilon_{k}
\end{equation}

Here, $q_{m}\simeq 1-(2R_{m})^{-1}$ indicates the fraction of
bounce particles in the total particles.

When we investigate one-half of a closed magnetic flux loop, we find
that the regime is very similar to that in an open magnetic flux
tube. It is reasonable to assume that the energetic particles will
be driven to move similar to what is shown in Figure 2. When
energetic particles leave the footpoint, they will be driven to move
upward. After this filtration (in fact, the filtration varies
continuously from the footpoint to the looptop), the area around the
looptop will be gathering energetic particles. When these energetic
particles reach a new thermal equilibrium by collision or
instability evolution, the temperature will become hotter than the
underlying atmosphere. Supposing a semicircle magnetic flux loop
$h=d/2$ and $d$=10$^{4}$ km, the magnetic field scale height near
the footpoint is about 1000 km. Then $R_{m}=3.75$,
$q_{m}\simeq0.87$, $\epsilon_{t0}\sim 9.5$ eV. The density of bounce
particles is about $N_{b}\simeq1.02\times10^{11}$ cm$^{-3}$ with
average thermal temperature of about $1.63\times10^{5}$ K. These
values are very similar to the observations of the loop with length
of about 5000 km. The dashed pink curve in Figure 3 shows the hot
region in the closed magnetic flux loop. Around the looptop, the
particle's longitudinal kinetic energy is much greater than its
transverse kinetic energy, and the temperature is also anisotropic
around the top region. As in the open magnetic flux tubes, the
temperature anisotropy will trigger plasma instabilities and lead to
secondary energy release and finally form a thermal isotropic
distribution (Dong et al. 1999).

In fact, there are various scales of closed magnetic flux loops in
the solar atmosphere. They may connect one side of a network
magnetic field to the other or connect an intranetwork magnetic
field to a different network magnetic field, and could connect one
polarity to another in sunspots, or even connect different active
regions (Lin 1995; Lin \& Rimmele 1999). The different scales of
closed magnetic flux loops extend to different heights in the solar
atmosphere. With the above MGP mechanism, the looptops will become
very hot. Figure 4 presents the schematic diagram of the closed
magnetic flux loop system with various scales. The thick parts with
pink or red colors show the hot region around the looptops. The
large amount of hot looptops may form the hot chromosphere and the
much hotter corona.

\begin{figure}[ht] 
\begin{center}
   \includegraphics[width=7.5 cm]{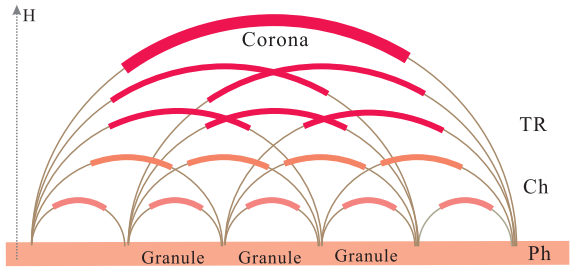}
\caption{Schematic diagram of the magnetic flux loops from the
photosphere (Ph) via the chromosphere (Ch) and the transition region
(TR) to the corona. The thick parts with pink or red colors show the
hot regions in the loops.}
\end{center}
\end{figure}

Additionally, when the energetic particles pile up and accumulate,
the density will increase as well as the temperature around the
looptops. Consequently, the plasma thermal pressure $p_{t}=nk_{B}T$
also increases. In an ideal situation, when the plasma pressure
exceeds the magnetic pressure $p_{m}=\frac{B^{2}}{2\mu_{0}}$
($\beta=\frac{p_{t}}{p_{m}}\geq 1$), the magnetic pressure cannot
balance the expanding trend of the plasma thermal pressure. Then the
magnetized plasma loop will break away from the confinement of the
magnetic field, disrupting it, and lead to a violent energy release.
From the balance between the plasma pressure and magnetic pressure
($p_{t}=p_{m}$), a density limit can be obtained:

\begin{equation}
n_{m}=\frac{B^{2}}{2\mu_{0}k_{B}T}
\end{equation}

For example, suppose that the looptop has a temperature of about 2
MK, and the magnetic field strength is about 50 Gs, then the density
limit is about 3.5$\times10^{11}$ cm$^{-3}$.

In fact, tokamak plasma experiments indicate that the $\beta$ limit
is always smaller than the unit ($\beta_{0}<1$, it means
$p_{t}<p_{m}$), and the density limit is also lower than that
obtained from Equation (14). The real value depends on the boundary
conditions (Haas \& Thomas 1973; Greenwald et al. 1988; Greenwald
2002), such as the radii of the magnetic loop and its cross-section,
the magnetic distribution in the section, etc. Generally, in recent
tokamak experiments, $\beta_{0}<10\%$. Therefore, the limit density
may be one order smaller than that deduced from Equation (14).

When the density of the hot plasma exceeds the limit, the energetic
particles may get free from the confinement of the magnetic field
and spread to the adjacent plasma. The final result leads to
annihilation of the loop (Inverarity \& Priest 1997) and energy
conversion between the energetic particles and the coronal plasmas.
The breakup around the hot looptop may be another formation mode of
solar eruptions in the chromosphere and corona. Observations
indicate that some solar flares begin to erupt at somewhere above
the plasma loops. It is possible that the cusp-like flares (Masuda
et al. 1995) may be formed from the breakup of hot looptops by the
MGP mechanism.

\section{Conclusions and Discussions}

The above analysis indicates that the MGP mechanism is a possible
model for explaining the formation of hot plasma upflows, such as
hot type II spicules observed by De Pontieu et al. (2009, 2011) and
the hot plasma ejections along the ultrafine magnetic loop channels
observed by Ji et al. (2012). It may provide another candidate to
answer the coronal heating problem. With this mechanism, we obtain
the following conclusions.

1. MGP mechanism can pick up the energetic charged particles from
the underlying atmosphere and move them upward, forming hot plasma
upflows, which pile up and accumulate in the upper chromosphere and
corona and finally cause the solar corona to become very hot.
Energetic particles come from the underlying atmosphere, which is
just the part of the high-energy tail in the thermal distribution
function of the plasmas. A preliminary estimation indicates that
this mechanism can cause the plasma temperature to reach about
10$^{4}$ K in the chromosphere, 10$^{5}$ K in the solar transition
region, and 10$^{6}$ K in the corona. The energy flow is estimated
to be about 8$\times10^{6}$ erg cm$^{-2}$s$^{-1}$ near the
photosphere, 6$\times10^{6}$ erg cm$^{-2}$s$^{-1}$ in the
chromosphere, 5$\times10^{6}$ erg cm$^{-2}$s$^{-1}$ in the
transition region, and 7$\times10^{5}$ erg cm$^{-2}$s$^{-1}$ in the
corona. Considering the inhomogeneous distribution of the magnetic
field in the solar atmosphere, the above estimations are compatible
with those in previous works (Withbroe \& Noyes 1977; Aschwanden et
al. 2007).

Additionally, the above estimations overlooked many important
effects, such as energy release by magnetic reconnection, various
wave damping, energy loss by radiation, and downflows from the
plasma cooling, etc. Magnetic reconnection takes place in some
small-scale regions with sheared magnetic configurations and wave
damping and resonance absorptions occur in twisting magnetic
configurations. Both are triggered by the underlying photospheric
convective motions. Their releasing energy increases our above
estimations around their action regions. At the same time, the
energy loss caused by the downward flows or radiation of ion-neutral
collisions in the partially ionized plasmas reduces the above
estimations to some extent in the lower region near the photosphere
and chromosphere. Different from these extremely dynamic processes,
MGP is a steady, continuous, and ubiquitous process occurring in
solar active regions, quiet regions, and in coronal holes. However,
as we lack enough knowledge about the magnetic field distributions
from the photosphere to the corona under present observations, it is
very difficult to make an exact assessment of the changes of the
above highly dynamic processes to our parameter estimations made by
the MGP mechanism.

2. As for the three key questions we mentioned in Section 1, the MGP
mechanism may present answers as follows. The energy heating the
solar chromosphere and corona comes from the solar interior; the
energetic charged particles are the energy carriers that are
transported in the open magnetic flux tubes or closed magnetic flux
loops by the MGP mechanism; the energetic particles deposit and
spread their energy through thermal collisions, plasma instability
triggered by temperature anisotropy, or the magnetic confinement
damage. The great number of open magnetic flux tubes and closed
magnetic flux loops with various space scales and considerable
magnetic gradients in the solar atmosphere make the solar corona
become hot. Additionally, as the upflow of the energetic charged
particles takes away a fraction of energy from the underlying
atmosphere, this will make the underlying atmosphere a bit cooler.
This fact can explain why the regions with strong magnetic fields
(such as in sunspot regions) near the solar surface have relatively
low temperatures than the adjacent regions.

3. In the MGP mechanism, the solar open magnetic flux tubes and
closed magnetic flux loops play a key role, acting as pumpers to
extract the energetic particles from the underlying thermal plasma,
and transport them to and deposit them in the upper atmosphere
(chromosphere and corona). This process creates a hot upflow with a
velocity of about 40 km s$^{-1}$ in the chromosphere and about 130
km s$^{-1}$ in the corona. The estimated velocities of upflows are
very close to the observations of type II spicules, which provides a
natural explanation of the formation of type II hot spicules and the
upward injections of hot plasma stretching from the photosphere to
the base of the solar corona.

However, the above estimations from the MGP mechanism are very
rough. Because the most important parameter in the above estimations
is the magnetic gradient ($\nabla B$), which dominates the final
results and will change continuously from the solar surface to the
upper corona and from the quiet region to the active region. So far,
however, we have no reliable measurement of the magnetic fields from
the chromosphere to the corona. Gelfreikh (1997) developed a method
for estimating the coronal magnetic field on the basis of
polarization inversion due to ordinary and extraordinary mode
coupling in the coronal region of quasi-transverse propagation of
radio observations. It can simultaneously provide the strength and
gradient of the coronal magnetic field from imaging observations
with high accuracy and high spatial resolution. When the new radio
telescope arrays such as the Chinese Spectral Radioheliograph (Yan
et al. 2009) and the American Frequency Agile Solar Radiotelescope
(Bastian et al. 2003) with broadband frequency and high
spectral-spatial resolutions, come into service, it will be possible
to obtain the three-dimensional magnetic maps of the solar
chromosphere and corona. Subsequently, a reasonable estimation of
the temperatures, density, and the velocity of upflows can be
obtained by the method proposed in this work.

The MGP mechanism, in its essence, implies that the inhomogeneous
magnetic fields coupling with plasmas will result in an
inhomogeneous distribution of plasma temperature. As magnetic fields
with a considerable gradient are ubiquitous in other stars, it is
possible to use the MGP mechanism to explain many phenomena
occurring in the stellar atmosphere such as the hot stellar corona
and some ejections. In astrophysical conditions, high speed jets are
observed frequently, these are called astrophysical jets (Meier et
al. 1997). In fact, the magnetic fields in the source regions
related to these jets are most similar to open flux tubes with
considerable magnetic gradients. Such a magnetic configuration can
naturally produce the hot upflows with high speed. For example, the
plasma temperature may exceed 10$^{7}$ degrees in the solar flaring
region and the adjacent atmosphere of black holes.

However, so far, coronal heating is still not resolved. The next
step in the work of the MGP model is to deduce the magnetic gradient
profile from the solar surface to the corona based on global MHD
models and new spectral imaging observations, then derive the
temperature profile and make a comprehensive comparison of this
profile to other observational results, such as the STEREO EUVI data
(Vazquez et al. 2010).

\acknowledgments The author thanks the referee for helpful and
valuable comments on this paper. Prof. J.Q. Dong and L.W. Yan
provided important suggestions. Dr. Y. Zhang was a great help in
improving the manuscript's language. This work is supported by NSFC
grants 11273030, 11373039, 11221063, 11433006, and MOST grant
2011CB811401.


\begin{thebibliography}{}

\bibitem[Altschuler et al.(1968)]{Altschuler1968}Altschuler, M.D., Lilliequist, C.G., \& Nakagawa, Y. 1968, SoPh, 5, 366

\bibitem[Aschwanden et al.(2007)]{Aschwanden2007}Aschwanden, M.J., Winebarger, A., Tsiklauri, D., \& Peter, H. 2007, ApJ, 659, 1673

\bibitem[Aschwanden(2005)]{Aschwanden2005}Aschwanden, M.J., \& Nightingale, R.W. 2005, ApJ, 633, 499

\bibitem[Bastian(2003)]{Bastian2003}Bastian, T.S. 2003, AdSpR, 32, 2705

\bibitem[Cargill(2004)]{Cargill2004}Cargill, P.J., \& Klimchuk, J.A. 2004, ApJ, 605, 911

\bibitem[Centeno et al.(2007)]{Centeno et al. 2007}Centeno, R., Socas-Navarro, H., Lites, B., et al. 2007, ApJ, 666, 137

\bibitem[Chen(2012)]{Chen2012}Chen, L., \& Wu, D.J. 2012, ApJ, 754, 123

\bibitem[Davila(1987)]{Davila1987}Davila, J. 1987, ApJ, 317, 514

\bibitem[De Pontieu et al.(2011)]{De Pontieu et al. 2011}De Pontieu, B., McIntosh, S.W., Carlsson, M., et al. 2011, Sci, 331, 55

\bibitem[De Pontieu et al.(2009)]{De Pontieu et al. 2009}De Pontieu, B., McIntosh, S.W., Hansteen, V.M., Schrijver, C. 2009, ApJL, 701, L1

\bibitem[De Pontieu(2007)]{De Pontieu2007}De Pontieu, B., McIntosh, S.W., Carlsson, M., et al. 2007, Sci, 318, 574

\bibitem[Dong(1999)]{Dong1999} Dong, J.Q., Chen, L., Zonca, F. 1999, NucFu, 39, 1041

\bibitem[Edlen(1943)]{Edlen 1943} Edlen, B. 1943, ZA, 22, 30

\bibitem[Filippov(2007)]{Filippov2007}Filippov, B., Koutchmy, S., \& Vilinga, J. 2007, A\&A, 464, 1119

\bibitem[Filippov(2009)]{Filippov2009}Filippov, B., Golub L., \& Koutchmy, S. 2009, SoPh, 254, 259

\bibitem[Fisk(2001)]{Fisk2001}Fisk, L.A., \& Schwadron, N.A. 2001, ApJ, 560, 425

\bibitem[Gelfreikh(1997)]{Gelfreikh1997}Gelfreikh, G.B., Pilyeva N.A., \& Ryabov, B.I. 1997, SoPh, 170, 253

\bibitem[Golub(1977)]{Golub1977}Golub, L., Krieger, A.S., Harvey, J.W., Vaiana, G.S. 1977, SoPh, 53, 111

\bibitem[Greenwald(2002)]{Greenwald2002}Greenwald, M. 2002, PPCF, 44, 27

\bibitem[Greenwald(1988)]{Greenwald1988}Greenwald, M., Terry, J.L., Wolfe, S.M., \& et al. 1988, NucFu, 28, 2199

\bibitem[Haas(1973)]{Haas1973}Haas, F.A., \& Thomas, C.L. 1973, PhFl, 16, 152

\bibitem[Habbal(1981)]{Habbal1981}Habbal, S.R., \& Withbroe, G.L. 1981, SoPh, 69, 77

\bibitem[Heyvaerts(1983)]{Heyvaerts1983}Heyvaerts, J., \& Priest, E.R. 1983, A\&A, 117, 220

\bibitem[Hagyard(1983)]{Hagyard1983}Hagyard, M.J., Teuker, D., Tandberg-Hanssen, E., et al. 1983, SoPh, 84, 13

\bibitem[Hahn(2013)]{Hahn2013}Hahn, M., \& Savin, D.W. 2013, ApJ, 763, 106

\bibitem[Hollweg(2006)]{Hollweg2006}Hollweg, J.V. 2006, JGRA, 111, A12106

\bibitem[Inverarity(1997)]{Inverarity1997}Inverarity, G.W., \& Priest, E.R. 1997, in ASP Conf. Ser. 111, Magnetic Reconnection
in the Solar Atmosphere, ed. R.D. Bentley \& J.T. Mariska (San
Francisco, CA:ASP), 296

\bibitem[Jess(2009)]{Jess2009} Jess, D.B., Mathioudakis, M., Erdelyi, R., et al. 2009, Sci, 323, 1582

\bibitem[Ji(2012)]{Ji 2012} Ji, H.S., Cao, W.D., \& Goode, P.R. 2012, ApJL, 750, L25

\bibitem[Klimchuk(2006)]{Klimchuk2006} Klimchuk, J. 2006, SoPh, 234, 41

\bibitem[Kerr(2012)]{Kerr2012} Kerr, R.A. 2012, Sci, 336, 1099

\bibitem[Landolfi(1987)]{Landolfi1987}Landolfi, M. 1987, SoPh, 109, 287

\bibitem[Lemen(2012)]{Lemen2012}Lemen, J. R., Title, A. M., Akin, D.J., et al. 2012, SoPh, 275, 17

\bibitem[Lin(1995)]{Lin1995} Lin, H.S. 1995, ApJ, 446, L421

\bibitem[Lin(1999)]{Lin1999} Lin, H.S., \& Rimmele, T. 1999, ApJL, 514, L448

\bibitem[Litwin(1993)]{Litwin1993} Litwin, C.; \& Rosner, R. 1993, ApJ, 421, 375

\bibitem[Liu(1996)]{Liu1996}Liu, Y., Wang, J.X., Yan, Y.H., \& Ai, G.X. 1996, SoPh, 169, 79

\bibitem[Lopez(2006)]{Lopez2006} L\'{o}pez Fuentes, M. C., Klimchuk, J. A., D\'{e}moulin, P. 2006, ApJ, 639, 459

\bibitem[Masuda(1995)]{Masuda1995} Masuda, S., Kosugi, T., Hara, H., Sakao, T., Shibata, K., \& Tsuneta, S. 1995, PASJ, 47, 677

\bibitem[Mathew(2000)]{Mathew2000}Mathew, S.K., \& Ambastha, A. 2000, SoPh, 197, 75

\bibitem[Meier(1997)]{Meier1997} Meier, D.L., Edgington, S., Godon, P., Payne, D.G., Lind, K.R. 1997, Natur, 388, 350

\bibitem[Narain(1996)]{Narain1996} Narain, U., \& Ulmschneider, P. 1996, SSRv, 75, 453

\bibitem[Parker(1988)]{Parker1988}Parker, E.N. 1988, ApJ, 330, 474

\bibitem[Parashar(2013)]{Parashar2013}Parashar, T.N., Velli, M., \& Goldstein, B.E. 2013, in AIP Conf. Proc. 1539, Solar Wind 13
ed. Gary, P. Z., Joe, B., Roberto, B., et al.(Melville, NY: AIP), 54

\bibitem[Priest(2002)]{Priest2002}Priest, E.R., Heyvaerts, J.F., \& Title A.M. 2002, ApJ, 576, 533

\bibitem[Pneuman(1983)]{Pneuman1983} Pneuman, G.W. 1983, ApJ, 265, 468

\bibitem[Pneuman(1984)]{Pneuman1984} Pneuman, G.W. 1984, SoPh, 94, 387

\bibitem[Rappazzo(2008)]{Rappazzo2008} Rappazzo, A.F., Velli, M., Einaudi, G., \& Dahlburg, R.B. 2008, ApJ, 677, 1348

\bibitem[Reid(2014)]{Reid2014} Reid, H.A.S., \& Ratcliffe, H. 2014, RAA, 14, 805

\bibitem[Schluter(1957)]{Schluter1957} Schluter, A., 1957, IAUS, 4, 356

\bibitem[Sturrock(1999)]{Sturrock1999} Sturrock, P.A. 1999, ApJ, 521, 451

\bibitem[Takakura(1970)]{Takakura1970} Takakura, T., \& Scalise, E. 1970, SoPh, 11, 434

\bibitem[Van Ballegooijen(2011)]{Van Ballegooijen2011}Van Ballegooijen, A.A., Asgari-Targhi, M., Cranmer, S.R., \& DeLuca, E.E. 2011, ApJ, 736, 3

\bibitem[Vasquez(2010)]{Vasquez2010} Vasquez, A.M., Frazin, R.A., \& Manchester IV, W.B. 2010, ApJ, 715, 1352

\bibitem[Verth(2011)]{Verth 2011} Verth, G., Goossens, M., \& He, J.S. 2011, ApJL, 733, L15

\bibitem[Vernazza(1981)]{Vernazza 1981} Vernazza, J.E., Avrett, E.H., \& Loeser, R. 1981, ApJS, 45, 635

\bibitem[Walsh(2003)]{Walsh2003}Walsh, R.W., \& Ireland, J. 2003, A\&ARv, 12, 1

\bibitem[Wedemeter(2012)]{Wedemeter2012}Wedemeter-B$\ddot{o}$hm, S., Scullion, E., Steiner, O., et al. 2012, Natur, 486, 505

\bibitem[Withbroe(1977)]{Withbroe1977}Withbroe, G.L., \& Noyes, R.W. 1977, ARA\&A, 15, 363

\bibitem[Wu(2003)]{Wu2003}Wu, D.J., \& Fang, C. 2003, ApJ, 596, 656

\bibitem[Yan et al(2009)]{Yan09}Yan, Y.H., Zhang, J., \& Wang, W., et al. 2009, EM\&P, 104, 97

\bibitem[Zhang(2011)]{Zhang 2011} Zhang, J., \& Liu, Y. 2011, ApJL, 741, L7

\bibitem[Zwaan(1978)]{Zwaan1978} Zwaan, C. 1978, SoPh, 60, 213

\end{thebibliography}
\end{document}